\documentclass[aps,twocolumn]{revtex4}
\usepackage{latexsym}
\usepackage{amsfonts}
\usepackage{graphicx} 
\usepackage{epsfig}
\usepackage{color}

\begin{document} 

\title{Thermodynamics and Dynamics of the Two-Scale\\ Spherically-Symmetric
Jagla Model of Anomalous Liquids}

\author{Limei Xu$^1$, Sergey V. Buldyrev$^{2,1}$, C. Austen Angell$^{3}$,
  H. Eugene Stanley$^1$}

\bigskip
\bigskip

\affiliation{$^1$Center for Polymer Studies and Department of Physics, Boston
  University,~Boston, MA 02215 USA\\ $^2$Department of Physics,~Yeshiva
  University, 500 West 185th Street,~New York, NY 10033 USA\\$^3$Department
  of Chemistry, Arizona State University, Tempe, Arizona 85287 USA }
  
\date{\today ~~ xbas.tex}


\begin{abstract}

Using molecular dynamics simulations, we study a liquid model which
consists of particles interacting via a spherically-symmetric two-scale
Jagla ramp potential with both repulsive and attractive ramps. The Jagla
potential displays anomalies similar to those found in liquid water,
namely expansion upon cooling and an increase of diffusivity upon
compression, as well as a liquid-liquid (LL) phase transition in the
region of the phase diagram accessible to simulations. The LL
coexistence line, unlike in tetrahedrally-coordinated liquids, has a
positive slope, because of the Clapeyron relation, corresponding to the
fact that the high density phase (HDL) is more ordered than low density
phase (LDL). When we cool the system at constant pressure above the
critical pressure, the hydrodynamic properties rapidly change from those
of LDL-like to those of HDL-like upon crossing the Widom line. The
temperature dependence of the diffusivity also changes rapidly in the
vicinity of the Widom line, namely the slope of the Arrhenius plot
sharply increases upon entering the HDL domain. The properties of the
glass transition are different in the two phases, suggesting that the
less ordered phase (LDL) is fragile, while the more ordered phase (HDL)
is strong, which is consistent with the behavior of
tetrahedrally-coordinated liquids such as water silica, silicon, and
BeF$_2$.

\end{abstract}

\maketitle

\section{Introduction}
An open question of general interest concerning liquid water is the relation
between a liquid-liquid (LL) phase transition and the dynamic properties
\cite{poole1, poole2, chenJCP2004, chen2005PC, PNAS,
Mallamace_PRL_2005,chen_PNAS_2006,Pradeepnature}. The LL phase transition may
have a strong effect on the dynamic properties of supercooled water,
including the glass transition \cite{Velikosci2001,JohariJCP2003}. In deeply
supercooled states, some glass-formers show ``strong'' behavior with a
well-defined activation energy, while other glass-formers display ``fragile''
behavior \cite{Angellsci1995}. Water appears to show a crossover between
fragile behavior at high T to strong behavior at low T
\cite{Angell_JPC_1993,Itonature1999, Bergman00, Poolenature2001,
Francisphysica2003}. The recent study on the Stillinger-Weber model of
silicon \cite{sastrynature2003}, which confirms the LL phase transition,
suggests that the less ordered high-density liquid (HDL) is fragile, while
the more ordered low-density liquid (LDL) is strong. These authors observed a
power-law singularity of the diffusivity in the less ordered HDL phase as it
approaches the spinodal of the LL transition at constant pressure. Recently
the fragility transition in nano-confined water was studied in neutron
scattering experiments pioneered by the Chen group at MIT
\cite{chenJCP2004,chen2005PC}, who found that water appears to show a
crossover between non-Arrhenius (``fragile'') behavior at high T to Arrhenius
(``strong'') behavior at low T \cite{Angell_JPC_1993,
Itonature1999,chenJCP2004,chen2005PC}. Their findings were confirmed using
NMR by Mallamace {\it et al.}  \cite{Mallamace_PRL_2005}.

A set of realistic water models---ST2, TIP5P, TIP4P, TIP3P, and
SPC/E---with progressively decreasing tetrahedrality which bracket the
behavior of real water have been studied to explore the generic
mechanisms of LL phase transition and anomalies associated with it
\cite{ poole1, poole2, PNAS, pgdbook, Mishima1998nature,
franzese2001nature, YamadaXX, pabloreview, lanaveprl, Paschek05, poole3,
pradeepPRE2005}. Studies \cite{franzese2001nature, Stell72,
gaussiancore,Stillinger97, Sadr98,Jagla99, Scala01, Buldyrev02} show
that tetrahedrality is not a necessary condition for anomalous behavior
and several spherically-symmetric potentials are indeed able to generate
density and/or diffusion anomalies.

In this article, we study a simplified model [Fig.~\ref{fig:jagla-pot}] based
on spherically-symmetric soft-core potentials with both attractive and
repulsive parts \cite{Jagla99}. Using extensive molecular dynamics (MD)
simulations, we study the static properties, and find that similar to water
models, the simple Jagla potential exhibits a density anomaly (TMD),
diffusivity anomaly as well as structural
anomaly~\cite{pablo_nature_2001}. We further explore the dynamic behavior
along different constant pressure paths in HDL (more ordered) and LDL (less
ordered) phase. We observe a non-Arrhenius behavior in the less ordered LDL
and Arrhenius behavior in the more ordered HDL phase, as well as a dynamic
crossover which occurs above the LL critical point as the system is cooled
down along constant pressure paths. We put forth a possible interpretation of
the dynamic crossover which occurs above the critical point in terms of the
LL critical point and its associated Widom line \cite{PNAS}. Thermodynamic
response functions -- constant pressure specific heat $C_{P}$, constant
temperature compressibility $K_{T}$ and the structural order parameters,
$Q_6$ (orientational) and $t$ (translational), support a relation between the
dynamic crossover and the LL phase transition.

The outline of the paper is as follows. In Sec. II, we introduce the
spherically-symmetric Jagla potential. In Sec. III, we describe the
method of the MD simulation. In Sec. IV, we define the quantities which
we study. In Sec. V, we investigate the static properties of the model,
while Sec. VI contains the simulation results of the dynamic
properties. Section VII discusses the relation with water and another
tetrahedral liquid BeF$_2$.

\section{Spherically-Symmetric Two-Scale Jagla Ramp Potential}

\begin{figure}[htb]
\includegraphics[width=6.5cm]{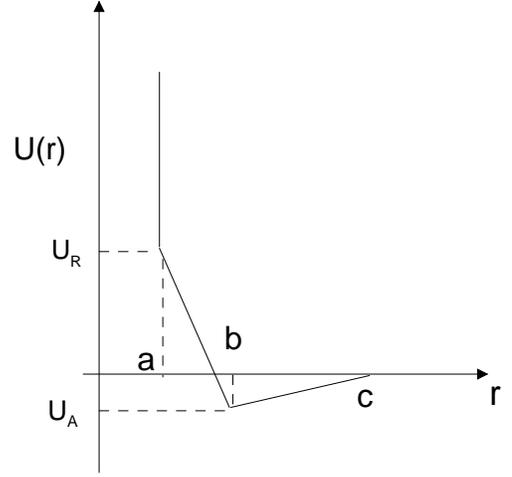}
\caption{(a) The spherically-symmetric ``two-scale'' Jagla ramp potential
with attractive and repulsive ramps. Here $U_{R}=3.5U_{0}$, $U_{A}=-U_{0}$,
$a$ is the hard core diameter, $b=1.72a$ is the soft core diameter, and
$c=3a$ is the long distance cutoff.}
\label{fig:jagla-pot}
\end{figure}

A spherically-symmetric potential with two different length scales has
been studied \cite{Sadr98,Scala01,Buldyrev02,Stell72,
Jagla99,franzese2001nature, PNAS}. Here, we study the linear ramp
potential but with both attractive and repulsive parts
\cite{Jagla99}. The potential is defined as
\begin{equation}
U(r) = \left\{
\begin{array}{ll}
\infty & r < a\\ U_A+(U_A-U_R)(r-b)/(b-a) & a<r<b,\\ U_A(c-r )/(c-b) &
b<r<c,\\ 0 & r > c
\end{array}\right.
\label{eq:potential}
\end{equation}
where $U_{R}=3.5U_{0}$ is the repulsive energy, $U_{A}=-U_{0}$ is the
attractive part, $a$ is the hardcore diameter, $b=1.72a$ is the well
minimum, and $c=3a$ is the cutoff at large distance
[Fig.~\ref{fig:jagla-pot}].

\section{Methods}

We apply the discrete MD method \cite{Buldyrev02, pradeepPRE}, approximating
the continuous potential Eq.~(\ref{eq:potential}) by step functions,
\begin{equation}
U_{n}(r) = \left\{
\begin{array}{ll}
\infty & r<a \\ U_{R}-k_1\Delta U_{1} & a + k_1\Delta r < r < a+(k_1+1)\Delta
r,\\U_{A} & b < r <b^{'},\\ U_{A} + k\Delta U_{2} & b^{'} + (k_2-1)\Delta
r^{'} < r < b^{'} + k_2\Delta r^{'},\\ 0 & r > c^{'}\end{array}\right.
\label{e-step}
\end{equation}
where $\Delta r\equiv 0.02a$, $\Delta r^{'}\equiv 0.16a$, $b^{'}=b+1/2\Delta
r^{'}$, $c^{'}=c-1/2\Delta r^{'}$, $\Delta U_{1}=(U_{R}-U_{0})/n_{1}$ with
$n_{1}=36$, $\Delta U_{2}=U_{0}/n_{2}$ with $n_{2}=8$, $0 \le k_1 \le
n_{1}-1$, and $ 1 \le k_2 \le n_{2}$ .

The standard discrete MD algorithm has been implemented for particles
interacting with step potentials \cite{rapaport, Buldyrev2003physicA,
sfa2002, franzese2001nature, Giancarlonature, pradeepPRE}. We use $a$ as the
unit of length, particle mass $m$ as the units of mass, and $U_{0}$ as the
unit of energy. The simulation time is therefore measured in units of
$a\sqrt{m/U_{0}}$, temperature in units of $U_{0}/k_{B}$, pressure in units
of $U_{0}/a^{3}$, and density $\rho\equiv Na^{3}/L^{3}$, where $L$ is the
size of the system and $N=1728$ is the number of particles. $NVE$, $NVT$ and
$NPT$ ensembles are applied in our simulation \cite{YamadaXX,
Buldyrev2003physicA}.

\section{The Four Quantities Studied}
 
\begin{itemize}

\item[{(a)}] The diffusion coefficient is defined as
\begin{equation}
D\equiv\lim_{t\to \infty} {\langle\left[{\bf \vec{r}}(t'+t)-{\bf \vec{r}}(t')\right]^2
\rangle _{t'} \over 6t},
\label{Dt}
\end{equation}
where $\langle\ldots\rangle_{t'}$ denotes an average over all particles
and over all $t'$.

\item[{(b)}] The static structure factor for wave vector $\vec{q}$ is
$S(\vec{q})=F(\vec{q},t=0)$, where $F(\vec{q},t)$ is the intermediate
scattering function defined as
\begin{equation}
  F(\vec{q},t)\equiv \langle\rho(\vec{q},t)\rho(-\vec{q},0)\rangle
\end{equation}
with the wave vector $q=2i\pi/L, i=1,2,\ldots$ and the Fourier transform
of the density function
\begin{equation}
\rho(\vec{q},t)\equiv\sum_{j}\exp[-i\vec{q}\cdot \vec{r}_{j}(t)].
\end{equation}
%


\item[{(c)}] The translational order parameter \cite{yan_PRL_2005,
pablo_nature_2001}
\begin{equation}
 t\equiv\int_{0}^{r_{c}} |g(r)-1|dr
\end{equation}
where $r$ is the radial distance, $g(r)$ is the pair correlation
function, and $r_{c}=L/2$ is the cutoff distance. A change in the
translational order parameter indicates a change in the structure of the
system. For uncorrelated systems, the interaction in the system is
short-ranged with $g(r)=1$, leading to $t=0$; for long-range correlated
systems, the modulations in $g(r)$ persist over large distances, causing
$t$ to grow.

\item[{(d)}] The orientational order parameter $Q$ characterizing the
average local order of the system \cite{yan_PRL_2005} is defined as
\begin{equation}
Q_\ell\equiv\left[\frac{4\pi}{2\ell+1}
\sum_{m=-\ell}^{m=\ell}|Y_{\ell,m}|^{2} \right]^{1/2},
\end{equation}
where $Y_{\ell,m}(\theta,\phi)$ is the spherical harmonic function with
$\theta$, $\phi$ are angles between the central particle and its 12
nearest neighbors, and $\ell=6$. In general, the value of $Q_{6}$
increases as the local order of a system increases, e.g, $Q_{6}$=0.574
for the fcc lattice and $Q_{6}=0.289$ for uncorrelated systems.

\end{itemize}

\begin{figure}[htb]
\includegraphics[width=8cm,angle=0]{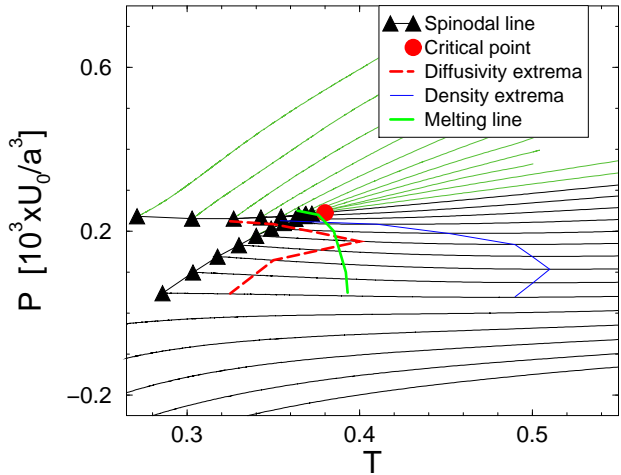}
\caption{The equation of state $P(T,\rho_{0})$ of the ramp potential for $26$
 values of $\rho_{0}\equiv Na^{3}/L^3$, where $L=15.0,15.2, 15.4,..., 20.0$
 is the cell edge. The LL critical point (closed circle) is located at
 $P=0.243$, and $T=0.37$, well above the equilibrium melting line (heavy solid
 curve). The gas-liquid critical point, not shown, is located at much higher
 temperature $T_{gl}=1.446$ ($P_{gl}=0.0417$ and $\rho_{gl}=0.102$). The
 density anomaly is represented by connecting all the points with $(\partial
 \rho/\partial T)_{P}=0$. The locus of the density extrema bounds the region
 of the diffusivity anomaly.}
\label{P-T}
\end{figure}

\begin{figure}[htb]
\includegraphics[width=8cm,angle=0]{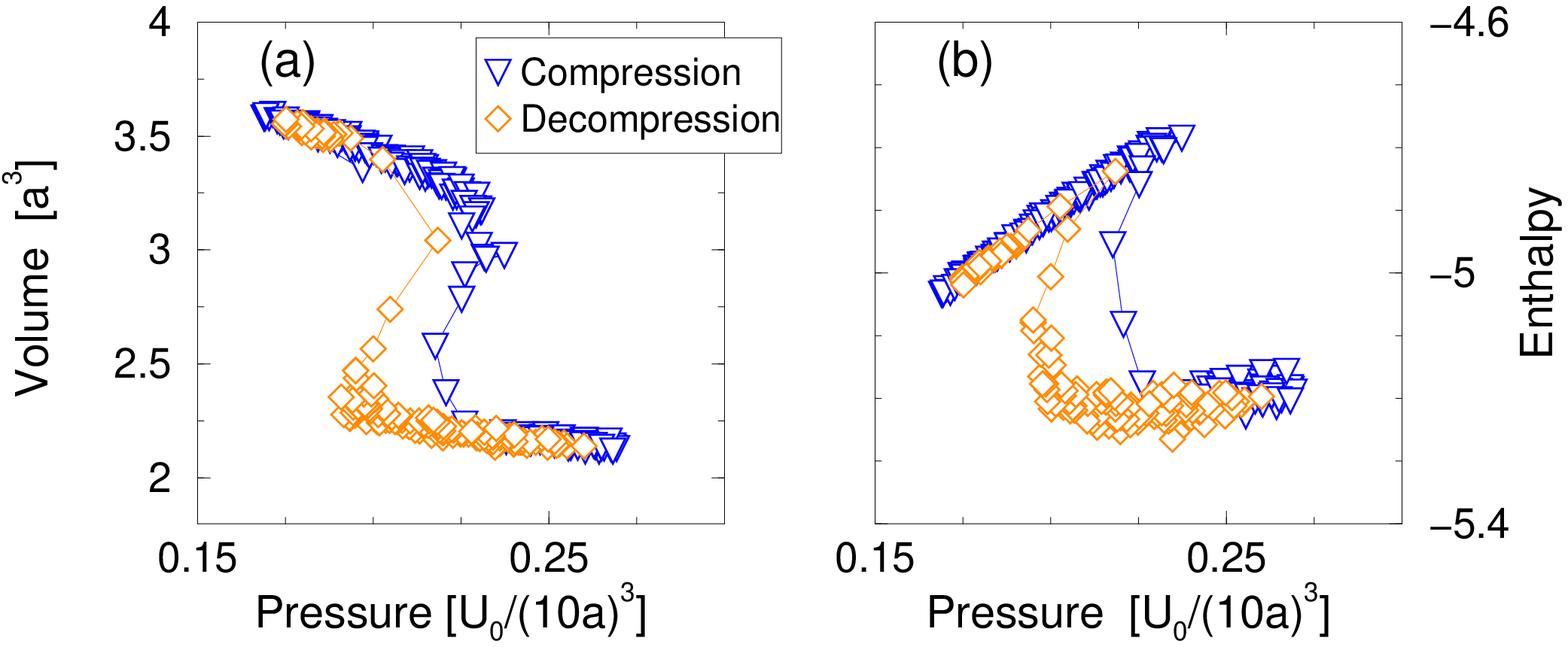}\hfill
\includegraphics[width=8cm,angle=0]{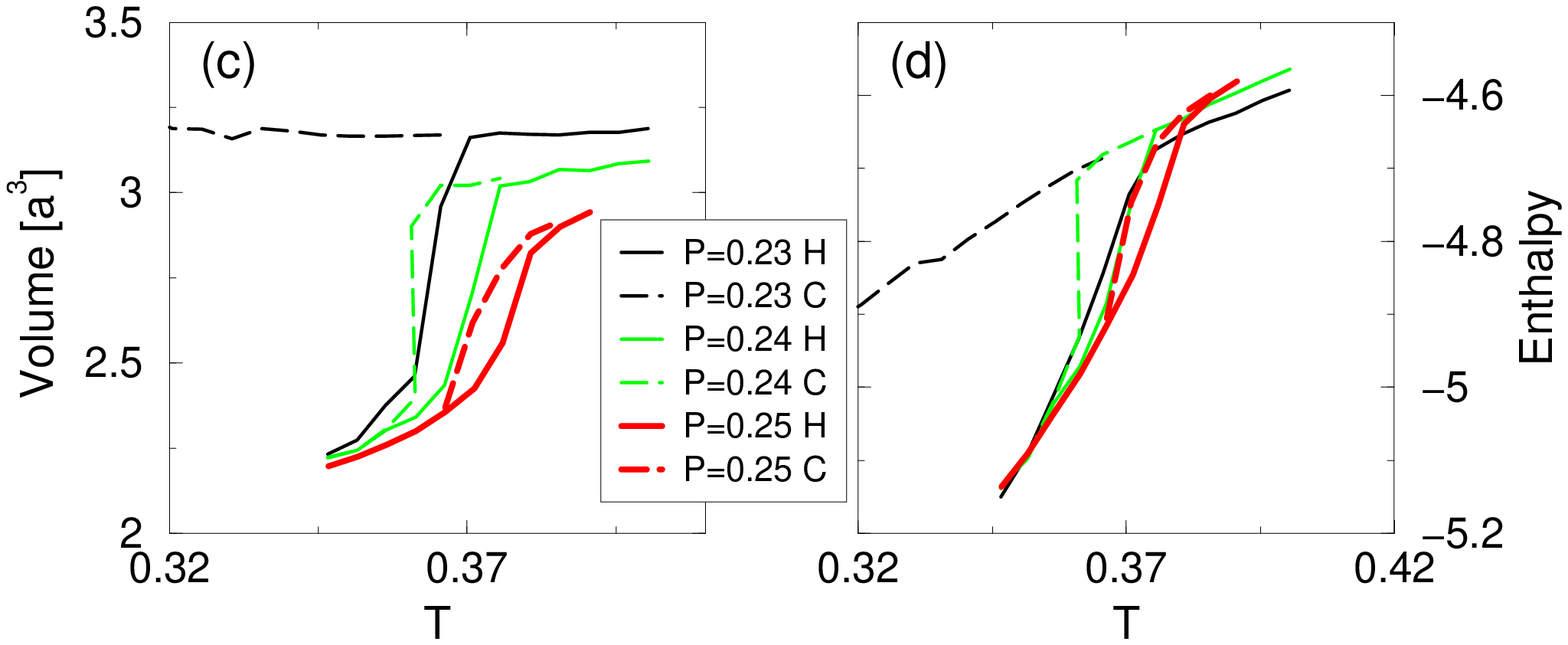}
\caption{Hysteresis upon compression and decompression along a constant
temperature path with $T=0.34$: (a) Volume per particle and (b) Enthalpy per
particle. Hysteresis upon cooling (C) and heating (H) along constant pressure
paths: (c) Volume per particle and (d) Enthalpy per particle. }
\label{enthalpy}
\end{figure}

\section{Simulation Results: Statics}

\subsection{Equation of state}

The equation of state of the Jagla model [Fig.~\ref{P-T}] is obtained
using two steps: (1) constant volume simulation upon slowly cooling
(NVT-ensemble), which allows us to obtain the equation of an isochore in
a single run; (2) constant energy (NVE-ensemble) simulation of
individual state points of a particular interest. The spinodals are
defined by the crossing of isochores $P(T,\rho)$ and
$P(T,\rho+\Delta\rho)$, where
\begin{equation}
\left({\partial P\over\partial\rho}\right)_{T}=0. 
\end{equation}
The hysteresis of volume~[Fig.~\ref{enthalpy}(a)] and enthalpy
[Fig.~\ref{enthalpy}(b)] upon compression and decompression along constant
temperature $T=0.34$, and the hysteresis of volume~[Fig.~\ref{enthalpy}(c)]
and enthalpy~~[Fig.~\ref{enthalpy}(d)] upon heating and cooling along
constant pressure paths are consistent with the existence of a LL phase
transition. The LL critical point, with $T_{c}=0.375, P_{c}=0.243$ and
$\rho_{c}=0.37$, is located at the maximal temperature on the spinodals. The
coexistence line, obtained by the Maxwell rule by integrating the isotherms,
has a positive slope of $0.96\pm 0.02k_{B}a^{-3}$.

According to the Clapeyron equation
\begin{equation}
\frac{dP}{dT}=\frac{\Delta S}{\Delta V},
\end{equation}
the entropy in the HDL phase is lower than the entropy in the LDL
phase. Hence, the HDL phase is more ordered than the LDL phase, which is the
opposite of the LL transition found in simulations for water \cite{poole2}
and silicon \cite{Poolenature2001}. 

\begin{figure}[htb]
\includegraphics[width=7cm,angle=0]{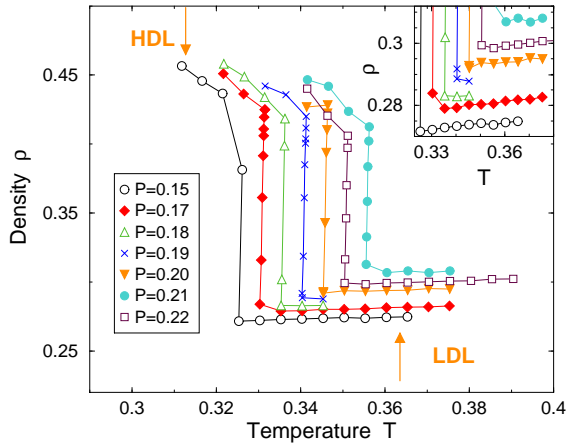}
\caption{Liquid -- liquid phase transition. Upon heating along constant
pressure paths below the critical point, the system experience a HDL to LDL
phase transition near the HDL spinodal line. The inset shows the density
anomaly upon heating along constant pressure paths.}
\label{HDL-LDL}
\end{figure}

We also studied the spontaneous liquid-liquid phase transition using $NPT$
ensemble [Fig.~\ref{HDL-LDL}]. Simulations of the isobaric heating of the HDL
liquid below the critical point show that the HDL phase loses its stability
and spontaneously changes into the LDL in the vicinity of the HDL spinodal
line [Fig.~\ref{HDL-LDL}]. The LL critical point of the
Jagla model, in contrast to those of water and silicon models, lies well
above the equilibrium melting line~[Fig.~\ref{P-T}] at which the solid and
liquid phases coexist and are in equilibrium.

Very recently, Gibson and Wilding \cite{Gibson_Arxiv_2006} study the
family of Jagla potentials with decreasing soft-core distance and found
that there is a parameter range within which the critical point lies
below the crystallization line, and the coexistence line has a negative
slope, resembling the situation for water.

\begin{figure}[htb]
\centerline{
\includegraphics[width=7cm,angle=0]{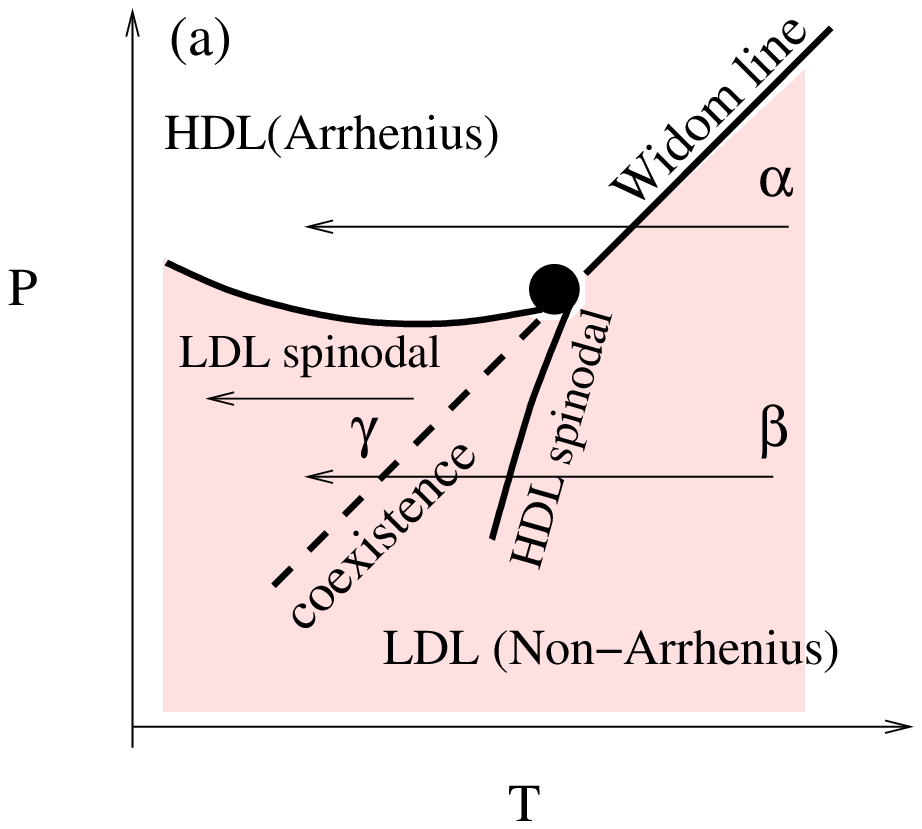}}
\centerline{
\includegraphics[width=7cm,angle=0]{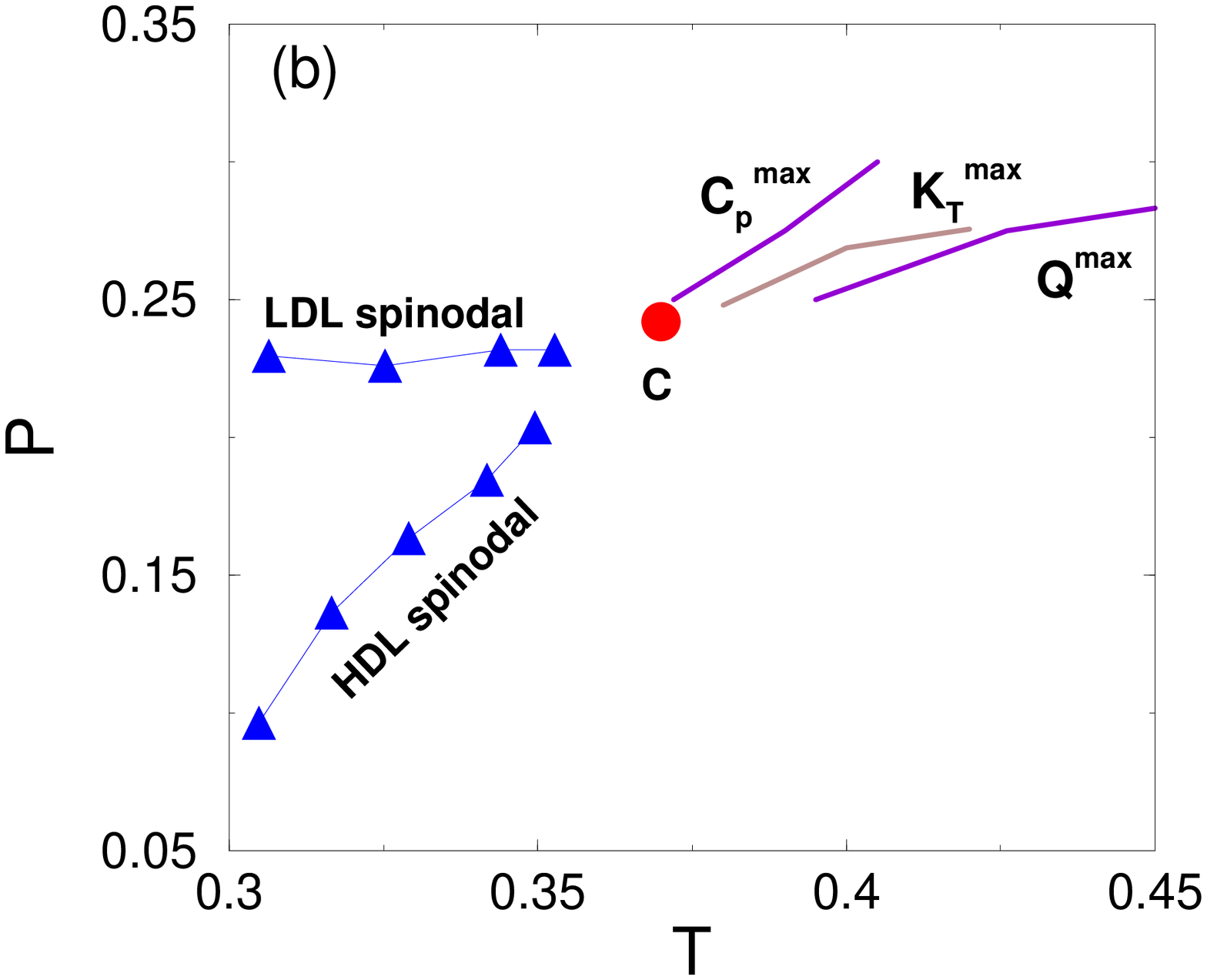}}
\caption{A sketch of the $P-T$ phase diagram for the two-scale Jagla
model. We study three different paths in the vicinity of the LL critical
point: (i) $P_{0}>P_{c}$ (path $\alpha$). Upon cooling along path $\alpha$,
the liquid changes from a low density state (characterized by a non-Arrhenius
dynamic behavior) to a high density state (characterized by Arrhenius dynamic
behavior) as the path crosses the Widom line. (ii) $P_{0}<P_{c}$ (path
$\beta$). Upon cooling along path $\beta$, the liquid remains in the LDL
phase as long as path $\beta$ does not cross the LDL spinodal line. Thus one
does not expect any dramatic change in the dynamic behavior along the path
$\beta$. (iii) $P_{0}<P_{c}$ (path $\gamma$). Upon cooling along path $\gamma$,
the liquid remains in HDL phase. Its dynamics, different from that of the LDL
phase, follows Arrhenius behavior. (b) The $P-T$ phase diagram for the
two-scale Jagla model. The loci of the specific heat maximum $C_{P}^{\rm
max}$, the compressibility maximum $K_{T}^{\rm max}$, and orientional order
parameter maximum $Q^{\rm max}$ are similar but not quite identical.}
\label{phasediagram}
\end{figure}
\subsection{Density Anomaly}

The temperature of maximum density (TMD) line is defined by $(\partial
V/\partial T)_{P}=0$. Due to the general thermodynamic relation
\begin{equation}
\left(\frac{\partial V}{\partial T}\right)_{P}=-\left(\frac{\partial
P}{\partial T}\right)_{V}\left(\frac{\partial V}{\partial P}\right)_{T},
\end{equation}
the TMD line coincides with the locus of points satisfying $(\partial
P/\partial T)_{V}=0$, which defines the pressure minimum on each
isochore (Fig.~\ref{P-T}). The density anomaly (density increase upon
heating along constant pressure paths) can also be seen in the inset of
Fig.~\ref{HDL-LDL}.

\begin{figure}[htb]
\centerline{
  \includegraphics[width=4.5cm]{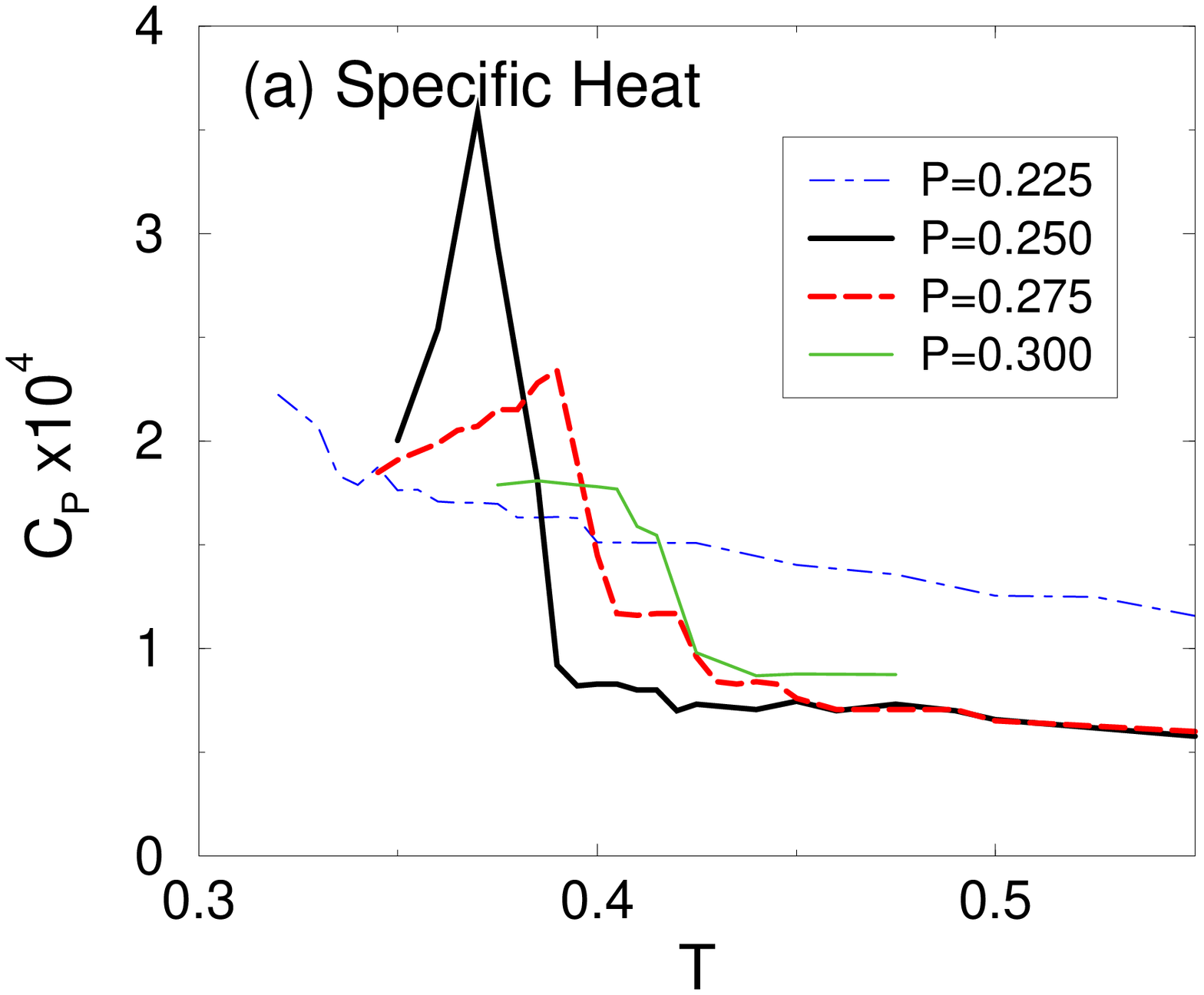}\hfill
  \includegraphics[width=4.5cm]{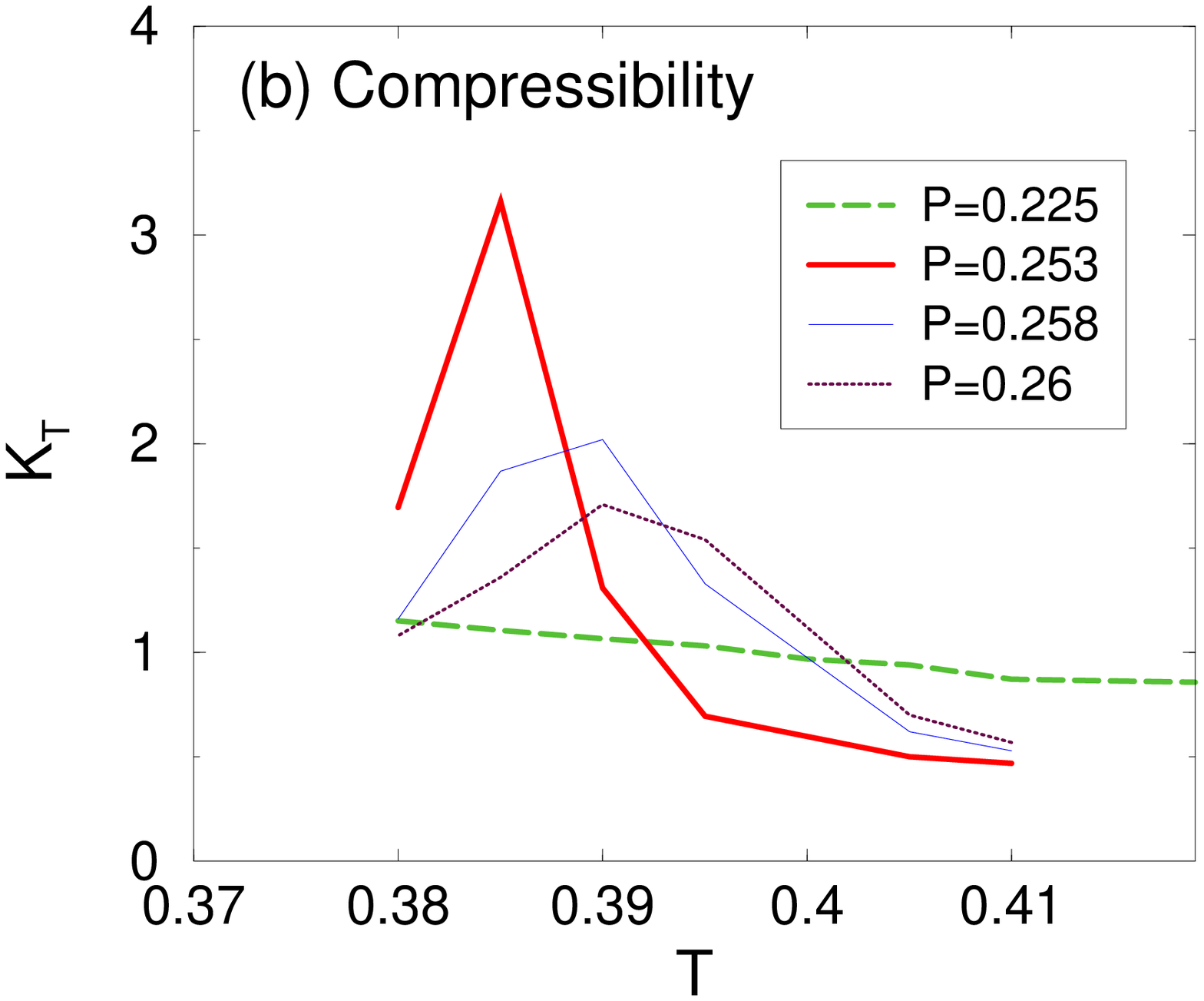}}
\caption{Response functions for the Jagla model as function of temperature
for different values of $P_{0}>P_{c}$~[Fig.~\ref{phasediagram}, path
$\alpha$]. (a) Constant pressure specific heat $C_{P}$ and (b) constant
temperature compressibility $K_{T}$. Both $C_{P}$ and $K_{T}$ have maxima, as
is known to occur experimentally \cite{Anisimovbook,Oguni}. For large $P_{0}$
the peaks become less pronounced and shift to higher temperature as the
system is farther away from the LL critical point.}
\label{Cp-Kt} 
\end{figure}

\subsection{Thermodynamics}
 
A sketch of the phase diagram based on the equation of state
[Fig.~\ref{P-T}] is shown in Fig.~\ref{phasediagram}(a) and the
simulated phase diagram is shown in
Fig.~\ref{phasediagram}(b). Different thermodynamic response functions
such as $C_P$ and $K_{T}$, which diverge at the critical point, have
maxima at temperatures $T_{\rm max}(P)$, which can be regarded as
temperatures on the extension of the coexistence line above the critical
temperature (the Widom line\cite{PNAS}) as we cool the system at
constant pressure $P_0>P_{c}$ \cite{Anisimovbook,AnisimovbookA,
AnisimovbookB, AnisimovbookC}.  We investigate $C_{P}$ and $K_{T}$ in
the Jagla model along paths $\alpha$ and path $\beta$
(Fig.~\ref{phasediagram}). We find that, along paths $\alpha$, $C_{P}$
has a maximum at $T_{\rm max}(P_{0})$ [Fig.~\ref{Cp-Kt}(a)], while $K_{T}$
has a maximum at a slightly different $T_{\rm max}(P_{0})$
[Fig.~\ref{Cp-Kt}(b)]. The response functions---$C_{P}$ and
$K_{T}$---increase continuously along path $\beta$
[Fig.~\ref{phasediagram}(a)] before the system reaches the stability
limit near the LDL spinodal \cite{pgdbook}.

\subsection{Structural order} 
\begin{figure}[htb]
\centerline{
 \includegraphics[width=4.3cm]{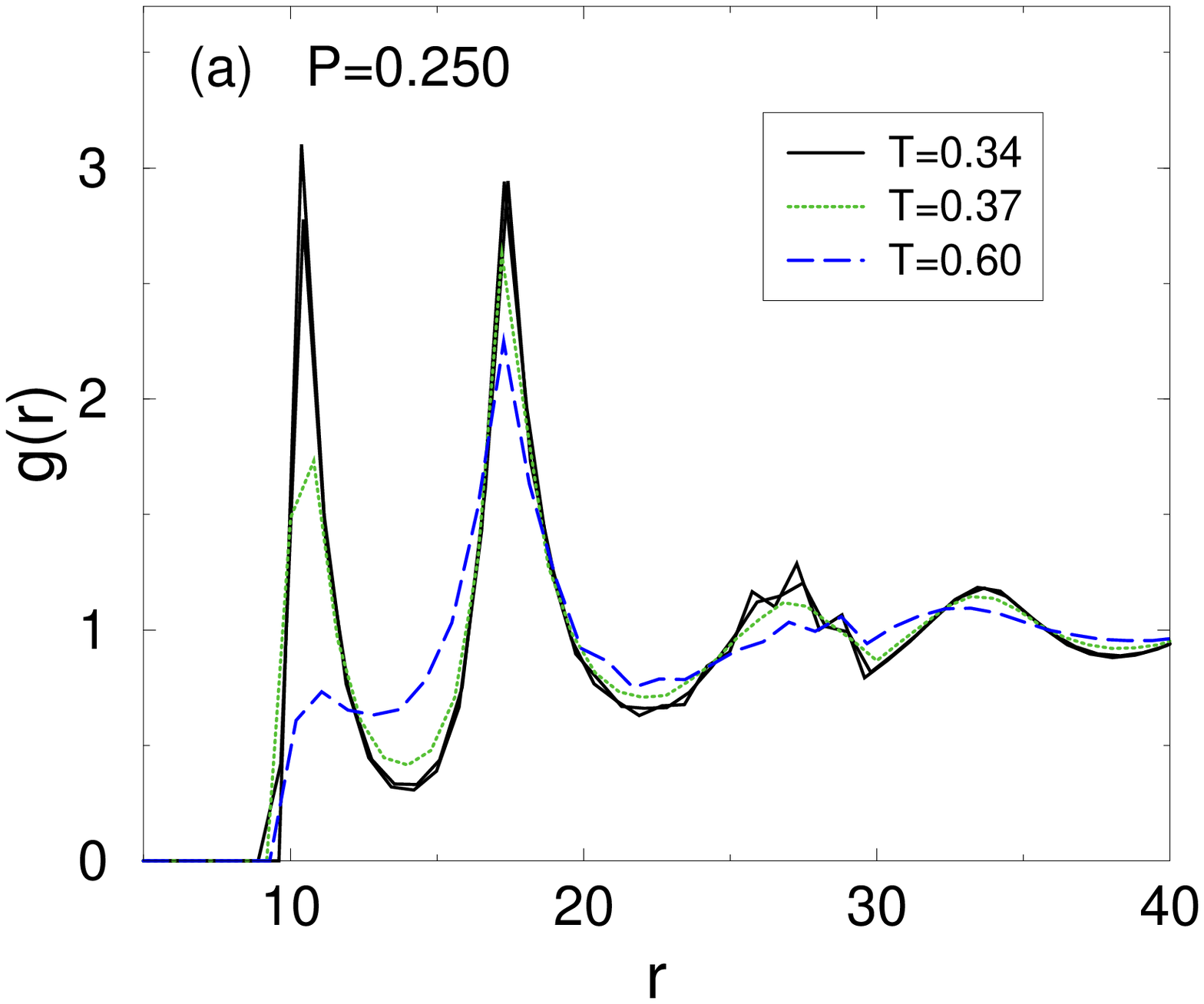}\hfill
 \includegraphics[width=4.3cm]{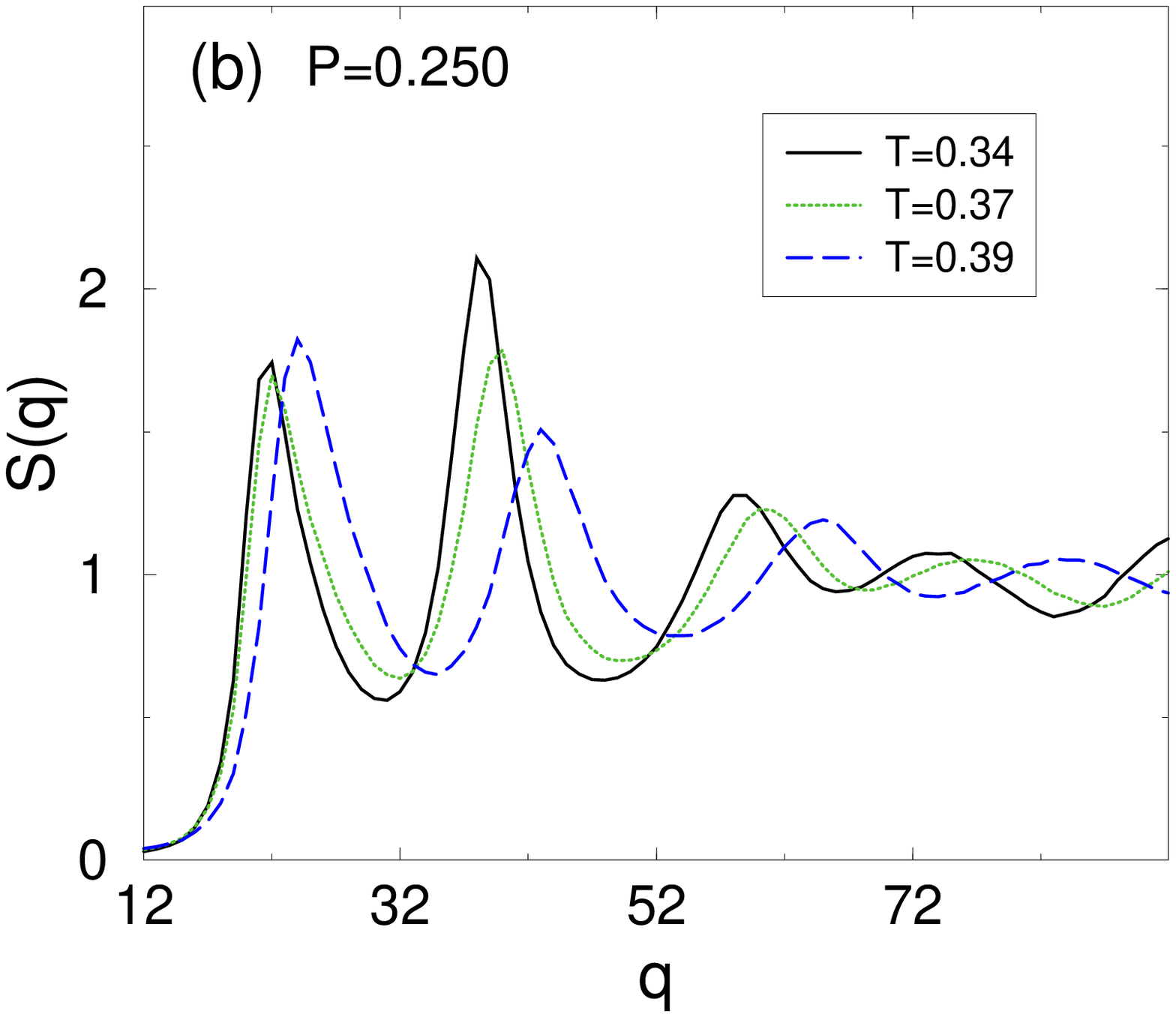}
}
\centerline{
 \includegraphics[width=4.3cm]{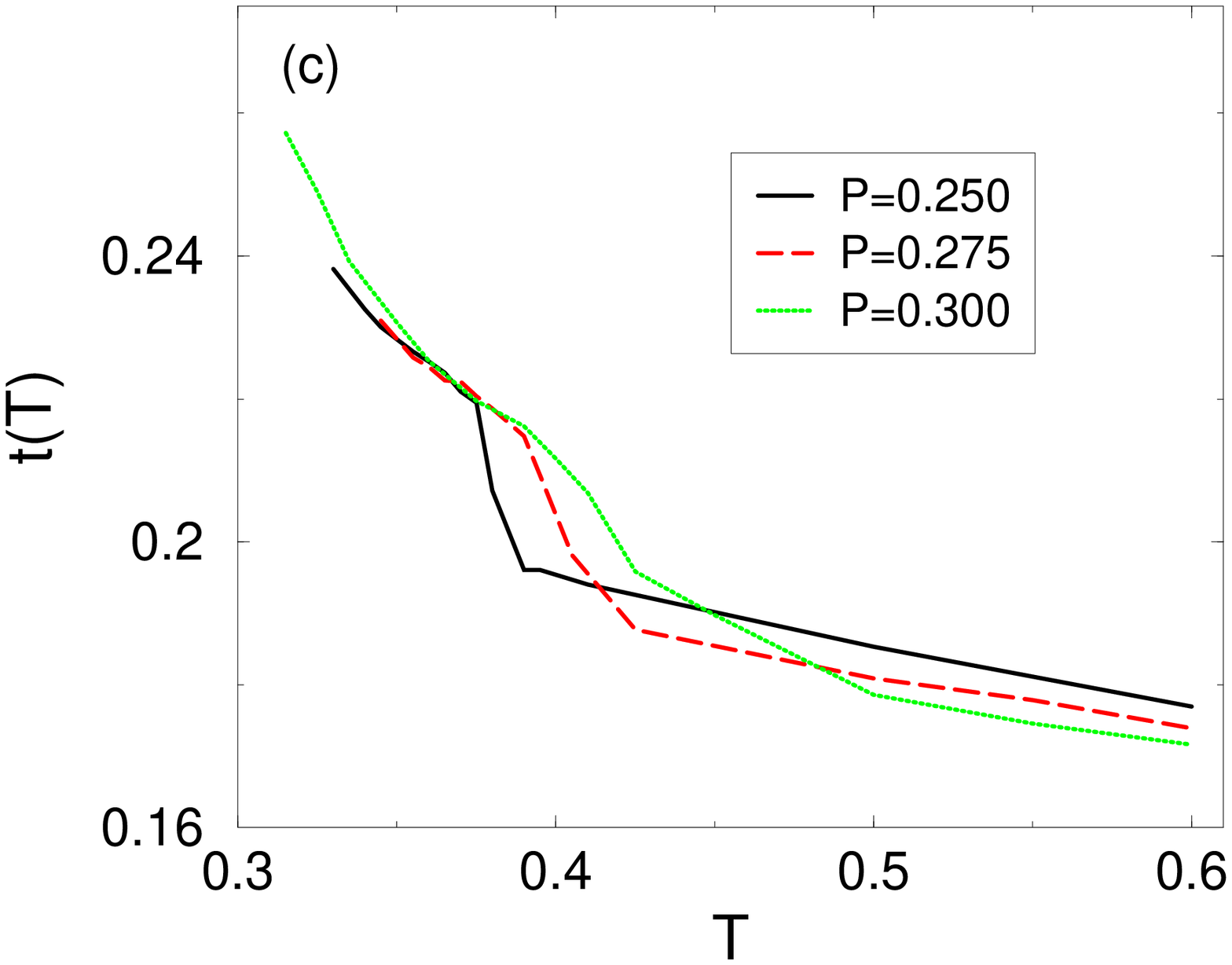}\hfill
 \includegraphics[width=4.3cm]{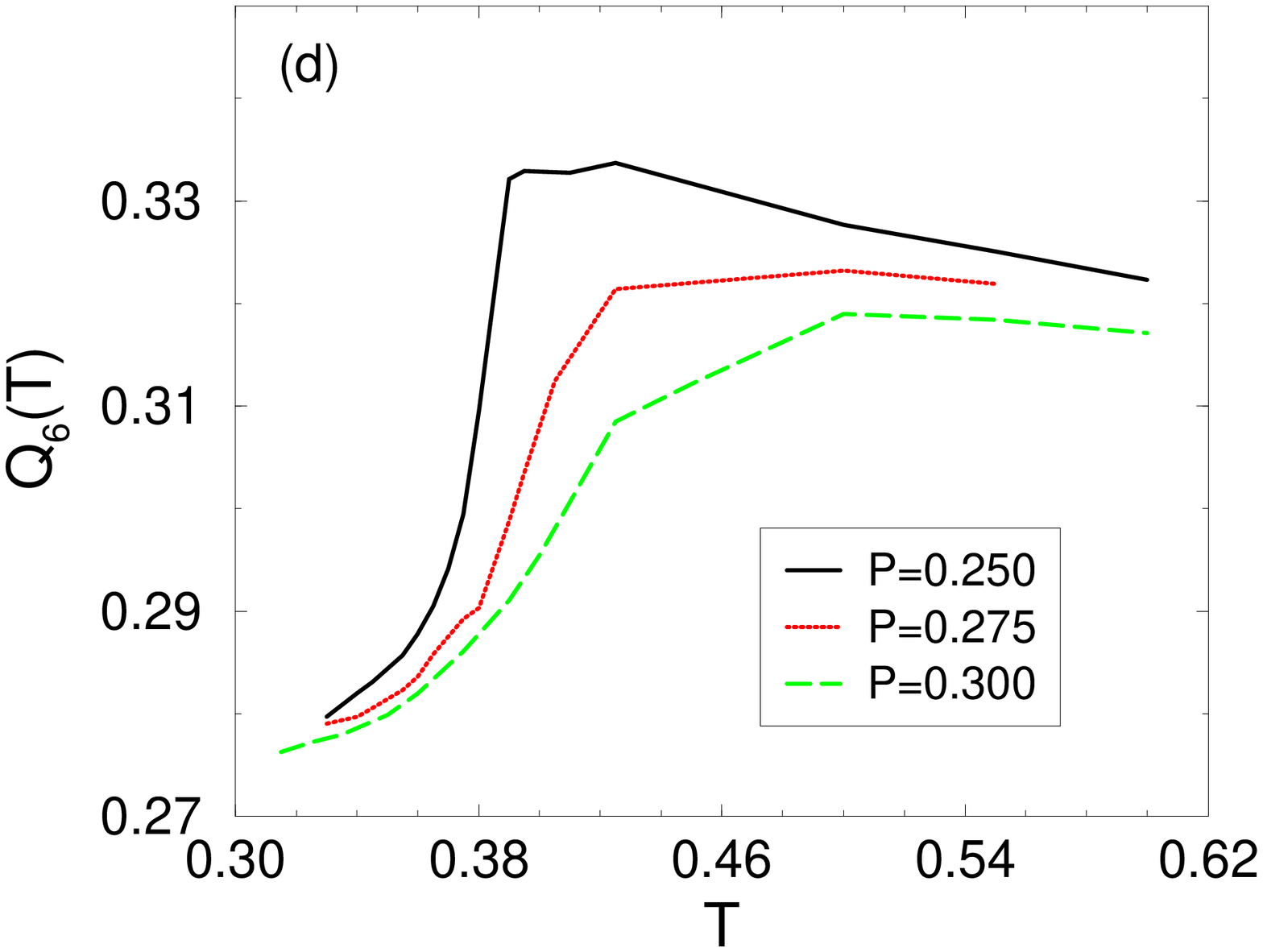}
}
\caption{(a) The pair correlation function $g(r)$ at constant pressure
$P_{0}=0.250$. The magnitude of the first peak indicates a HDL-like liquid at
low $T$ and LDL-like liquid at high $T$. (b) Distribution of $q$ vectors at
constant pressure $P_{0}=0.250$. The shifts of the first and second peaks in
the distribution of the $q$ vectors further indicates that the liquid changes
smoothly from HDL-like to LDL-like as it crosses the Widom line. (c) The
translational order parameter $t$ and (d) the orientational order
parameter $Q_{6}$ along constant pressure paths.}
\label{tT-fq} 
\end{figure}

Analogously, we can expect that the structural properties of the system
also change from those resembling the LDL phase to those resembling the
HDL phase when the system crosses the Widom line (the extension of the
coexistence line). The pair correlation function $g(r)$ for different
$T$ along a constant pressure path is shown in Fig.~\ref{tT-fq}(a). At
low temperature $g(r)$ exhibits a more pronounced first peak near the
hard core distance and a less pronounced peak at high $T$, indicating a
change from the LDL-like structure to HDL-like structure upon cooling
along paths $\alpha$. The same behavior can also be seen from the static
structure factor $S(q)$ [Fig.~\ref{tT-fq}(b)]. The sharp transition in
the translational order parameter $t$ [Fig.~\ref{tT-fq}(c)] and the
peaks in the orientational order parameter $Q_{6}$
[Fig.~\ref{tT-fq}(d)] above the LL critical point along path $\alpha$,
indicate that as the system crosses the Widom line region
[Fig.~\ref{phasediagram}(b)], the local structure of the system also
changes from LDL-like side to HDL-like side, which is consistent with
the features observed in the thermodynamic response functions.

\section{Simulation Results: Dynamics}

\subsection{Diffusivity Anomaly}

\begin{figure}[htb]
\includegraphics[width=7cm,angle=0]{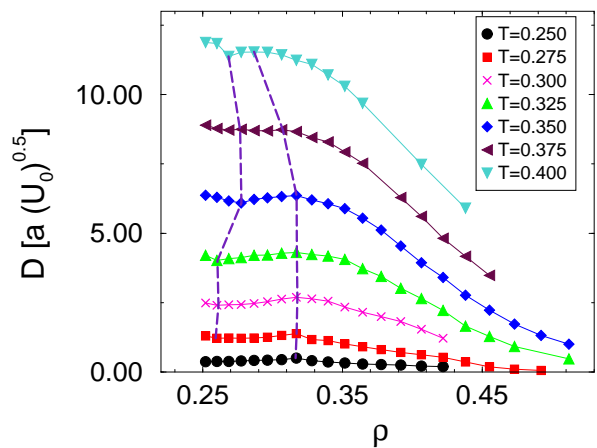}
\caption{Diffusivity as a function of density for seven values of 
  temperature. The diffusivity anomaly region, where $D$ increases with
  compression (density), is located within the diffusivity extrema lines
  (heavy dashed line).}
\label{DM}
\end{figure}

Figure~\ref{DM} shows the diffusivity $D(\rho)$ along seven
isotherms. There exists a diffusivity anomaly region along each isotherm
where the diffusivity {\it increases\/} upon compressing instead of
decreasing. The loci of the diffusivity extrema where $(\partial
D/\partial P)_{T}=0$ [heavy dashed lines in Fig.~\ref{P-T} and
Fig.~\ref{DM}] defines the diffusivity anomaly region.

We study the $T$ dependence of $D$ along three different constant
pressure paths [Fig.~\ref{phasediagram}(a)]:

\begin{itemize}

\item[{(i)}] Path $\alpha$, $P_0>P_{c}$ (one-phase region);

\item[{(ii)}] Path $\beta$, $P_0<P_{c}$ (in the LDL phase); and

\item[{(iii)}] Path $\gamma$, $P_0<P_{c}$ (in the HDL phase).
 
\end{itemize}

\begin{figure}[htb]
\centerline{\includegraphics[width=7cm]{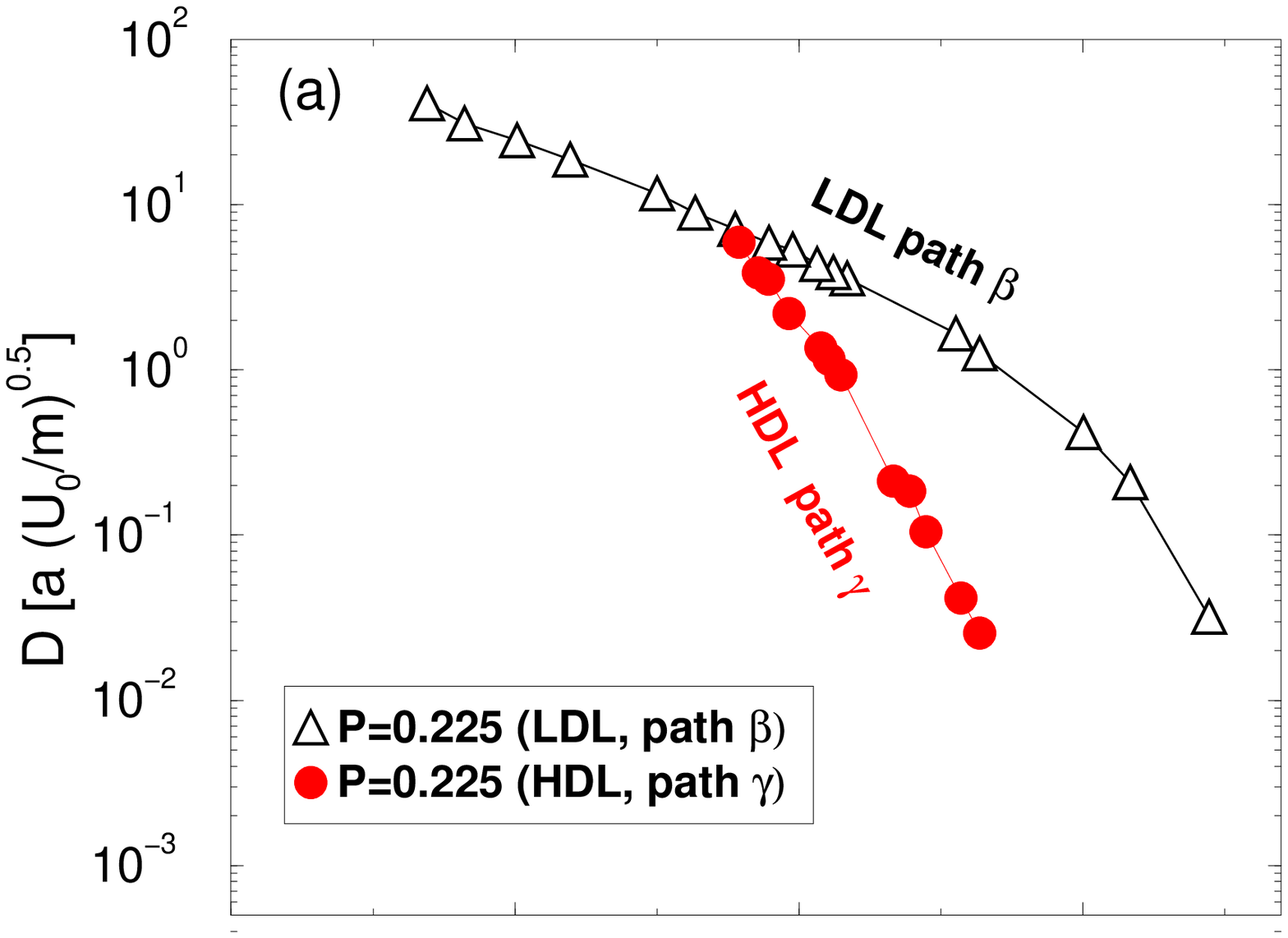}}
\centerline{\includegraphics[width=7cm]{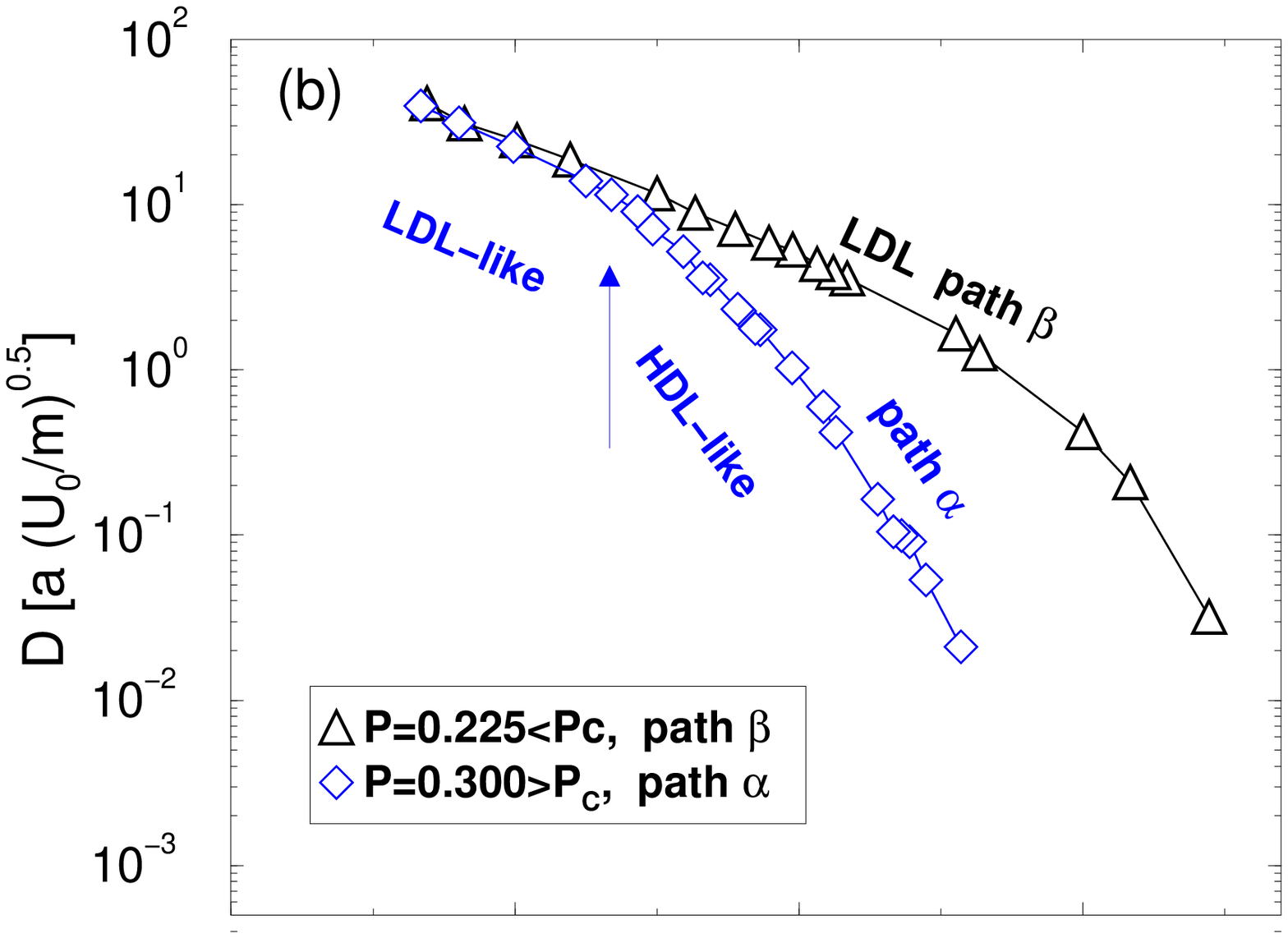}}
\centerline{\includegraphics[width=7cm]{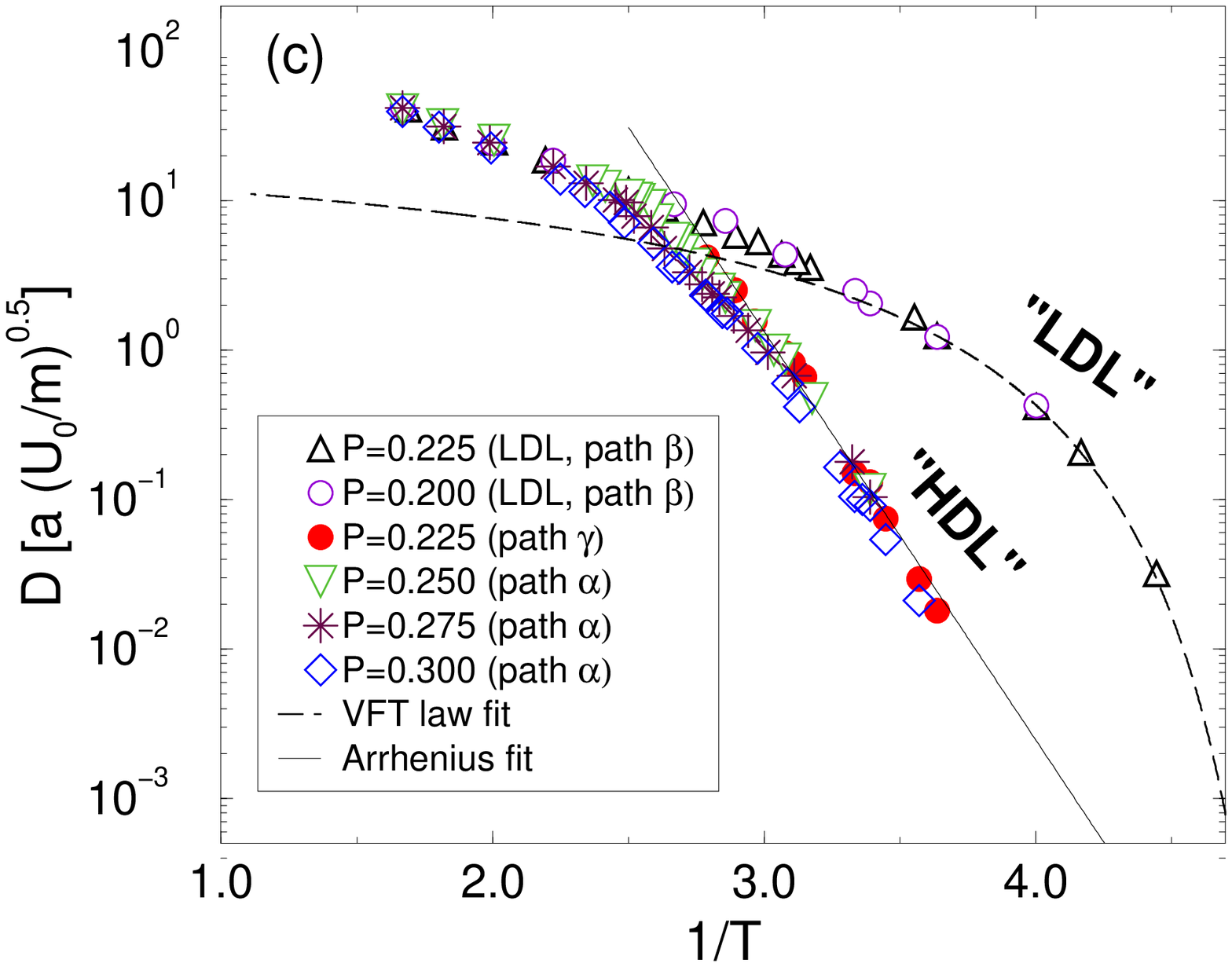}}
\caption{Dynamic behavior for Jagla potential. The $T$-dependence of
diffusivity $D$ along constant pressure paths: (a) $P_{0}=0.225<P_{c}$ for
both path $\beta$ and path $\gamma$. The more ordered phase (HDL) is strong,
while the less ordered phase (LDL) is fragile. (b) Path $\beta$ with
$P_{0}=0.225<P_{c}$, and path $\alpha$ with $P_{0}=0.300>P_{c}$. A dynamic
crossover occurs along constant pressure paths above the critical pressure
when the Widom line is crossed [Fig.~\ref{Cp-Kt}]. (c) Path $\beta$ with
$P_{0}=0.200, P_{0}=0.225<P_{c}$, path $\gamma$ with $P_{0}=0.225<P_{c}$, and
path $\alpha$ with $P_{0}=0.250, 0.300>P_{c}$~[Fig.~\ref{phasediagram}].}
\label{D-T-P}
\end{figure}
\subsection{Dynamics for $P<P_c$}

For $P_0<P_{c}$, the diffusivity $D$ at high temperature $T>T_{c}$
follows the Arrhenius law
\begin{equation}
\label{non-Arrhenius}
D=D_{0}\exp\left(-{E_A\over k_BT}\right)
\end{equation}
with a roughly pressure independent activation energy $E_A\approx
1.53$, while at low temperatures in the two-phase region, $D$ behaves
differently along path $\beta$ and path $\gamma$ upon cooling at
constant pressure [Fig.~\ref{D-T-P}]. Along path $\beta$
~[Fig.~\ref{phasediagram}(a)] which belongs to the LDL phase, the $T$
dependence of $D$ is non-Arrhenius and follows the Vogel-Fulcher-Tamann
law
\begin{equation} 
\label{Arrhenius}
  D=D_{0}\exp\left(-\frac{B}{T-T_{0}}\right)\sim
  D_{0}\exp\left(-\frac{T_{0}}{T-T_{0}}\frac{1}{f}\right).
\end{equation}
For $P=0.225$ the fitting parameters are $B\approx 0.2$, $T_{0}\approx
0.184$, and the fragility parameter $f=T_{0}/B\approx 0.66$
[Fig.~\ref{D-T-P}]. Quantitatively, the ``steepness index of the
fragility''~\cite{angell_JCP_1993} $m\approx 368$,
with the glass temperature $T_{g}\approx 0.192$ estimated as the temperature
at which $\exp(B/(T-T_{0}))=10^k$, where $k=16$ is a typical value for
thermally-activated systems \cite{angell_JCP_1993}. The large value of $m$
indicates that the behavior in the LDL phase resembles that of a very fragile
liquid. Note that the value of the index number $m$ largely
depends on the Vogel-Fulcher-Tamman fitting at the last data
point~[Fig.~\ref{D-T-P}(a)].

On the other hand, along paths $\gamma$ which belong to the HDL phase,
$D$ follows Arrhenius behavior with $E_{a}\approx 6.30 $, which is much
larger than the activation energy at high temperature.

\subsection{Dynamics for $P>P_c$}

For $P_0>P_{c}$ along path $\alpha$, there is a crossover in the
behavior of $D$ near $T\approx 0.4$
[Figs.~\ref{D-T-P}(b) and \ref{D-T-P}(c)].
Such a crossover, where $C_P$ passes through a maximum
[Fig.~\ref{Cp-Kt}(a)], would be predicted by the Adam-Gibbs equation
\cite{Francisphysica2003,AG},
\begin{equation}
D=D\exp\left(-\frac{C}{TS_{c}}\right),
\end{equation}
because of the rapid change in the temperature dependence on $S_{c}$ as
the Widom line is crossed. When $C_P$ has fallen to a smaller and more
slowly changing value, the temperature dependence of $D$ assumes an
Arrhenius behavior but with a somewhat larger slope than at high
temperatures where Fig.~\ref{Cp-Kt}(a) shows $C_P$ to be very small. The
behavior is similar to what was observed in experimental studies of the
strong liquid BeF$_2$ \cite{Angell_BeF2_2001} and in simulations of
SiO$_2$ \cite{Poolenature2001}. In fact, the parallel with the case of
BeF$_2$ \cite{Angell_BeF2_2001} is remarkable. In each case the
Arrhenius slope extrapolates to an intercept at $1/T=0$, which is six
orders of magnitude above the intercept of the high temperature
Arrhenius part of the plot (which is common to all phases). Thus, the
behavior of the HDL-like liquid on the low-temperature side of the Widom
line can be classified as that of a strong liquid. The behavior on the
high-temperature side of the Widom line, in the LDL-like phase, however,
is very different, resembling that of the fragile liquid (LDL-like), as
is clear from Figs.~\ref{D-T-P}(b) and \ref{D-T-P}(c). Thus, the present
spherically-symmetric Jagla ramp potential exhibits a dynamic crossover
from LDL-like (fragile liquid) at high-temperature to HDL-like (strong
liquid) at low-temperature, suggesting the analogous fragile-to-strong
transition as in water, with the difference that the strong liquid is
now the HDL phase.

\section{Discussion}

\subsection{Jagla Ramp Potential}

The mechanisms underlying the different dynamic behaviors we find can be
related to the LL phase transition [Fig.~\ref{phasediagram}(a)]. The
coexistence line has a positive slope, so we have one phase for $P>P_c$
and two phases---LDL and HDL---for $P<P_c$. According to the Clapeyron
equation, HDL entropy is lower than LDL entropy, so HDL is more ordered
than LDL which is the opposite of water. In the region of the P-T phase
diagram between the LDL and HDL spinodals, the system can exist in both
the LDL and HDL phases, one stable and one metastable.

\begin{itemize}

\item[{(i)}] The limit of stability of the less-ordered LDL phase is
determined by the high pressure LDL spinodal $P_{\rm LDL}(T)$, which, for our
model, is unlikely to be crossed by cooling the system at constant
pressure~[Figs.~\ref{enthalpy}(c) and (d)]---because $P_{\rm LDL}(T)\approx
P_c$ for all $T$ except in the immediate vicinity of the liquid-liquid
critical point~[Fig.~\ref{phasediagram}]. The dynamic behavior of the less
ordered LDL phase is non-Arrhenius, which is the characteristic of fragile
glass formers [Fig.~\ref{phasediagram}].

\item[{(ii)}] On the other hand, the limit of stability of the more ordered
HDL phase is determined by the low pressure HDL spinodal $T_{\rm HDL}(P)$
[Fig.~\ref{phasediagram}], which can be crossed by heating the HDL phase at
constant pressure~[Fig.~\ref{enthalpy}(c) and (d)]. That is why the dynamic
behavior of the more ordered HDL phase can be studied only when $T<T_{\rm
HDL}(P)$ for $P<P_{c}$.

\end{itemize}

\subsection{Comparison with Water}

\begin{figure}[htb]
\centerline{\includegraphics[width=7cm]{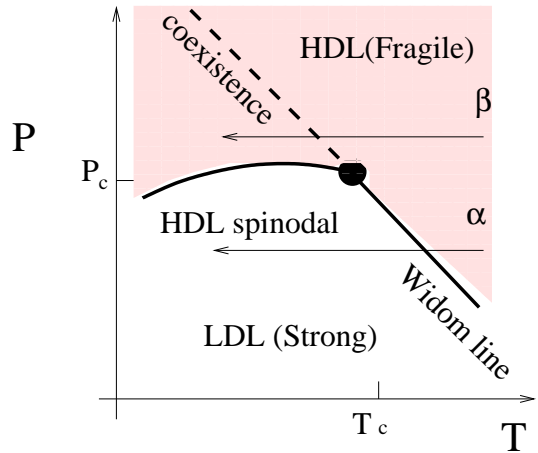}}
\caption{A sketch of the $P-T$ phase diagram for water models, showing
path$\alpha$ and $\beta$. Note that unlike the Jagla two-scale ramp potential
[Fig.~\ref{phasediagram}(a)], the HDL is less ordered than the LDL, so by the
Clapeyron relation, the Widom line has negative slope.}
\label{water}
\end{figure}

For the Jagla model, the dynamic crossover upon cooling for $P_0>P_c$ (Path
$\alpha$) is the same that was observed in realistic water models
\cite{Itonature1999,chenJCP2004}, which is a fragile to strong transition. In
water models which have a negatively sloped coexistence, the density of the
high temperature phase is larger than the density of the low temperature
phase~[Fig.~\ref{water}]. The density of the high temperature phase is larger
than the density of the low temperature phase if the coexistence line has a
negative slope, so the high temperature less ordered phase is the HDL
phase. When water is cooled at $P_0>P_c$ [Fig.~\ref{water}, path $\beta$], it
crosses the coexistence line but does not cross the spinodal of the less
ordered HDL phase, which may become almost horizontal line on the $P-T$ plane
as recent $ST2$ simulations suggest \cite{poole2}. Thus, water remains in the
metastable HDL phase even at very low temperatures, since the low-density
phase may not nucleate out of the HDL phase. Accordingly, above the critical
pressure, water may remain fragile even at very low temperatures. In
contrast,upon cooling at $P_{0}<P_{c}$ [Fig.~\ref{water}, path $\alpha$],
water crosses the Widom line, and its thermodynamic properties continuously
change from those of the HDL phase to those of the LDL phase. This is
demonstrated by the $C_P$ peak found by cooling water in small pores at
atmospheric pressure \cite{Oguni}. Therefore, for $P_0<P_c$ one expects a
fragile-to-strong transition upon cooling, which is indeed experimentally
observed \cite{chenJCP2004,chen2005PC}.
 
\subsection{Comparison with the Tetrahedral Liquid BeF$_2$}

Interestingly, what was observed in a combined MD/experimental study of
BeF$_2$~\cite{Angell_BeF2_2001,Moynihan, Nemilov} show a dynamic crossover
similar to what we observed for Jagla model for $P_0>P_{c}$
[Fig.~\ref{D-T-P}(a)]. In addition, the MD simulations of BeF$_2$ show a
density anomaly and specific heat maximum close to the point of the dynamic
crossover at about $2T_{g}$. In the Jagla model, the dynamic crossover and
the $C_P$ maximum occur at higher temperatures $T\approx 3.5 T_{g}$. The
difference between BeF$_2$ and the Jagla model is that a second LL critical
point has not been directly observed for BeF$_2$. Therefore, we can not call
the region of fast change of the dynamic and thermodynamic properties a Widom
line. However, the extrapolation of the simulation isochores in the density
anomaly region suggests possible existence of a critical point at lower
temperature and higher pressure. Thus,
the region of fast changes of the thermodynamic response functions is
possiblely associated with a Widom line emanating from the hypothetical
critical point \cite{Brazhkin_book_2002,Brazhkin_book_2002b} hidden in the
inaccessible region. As in water, this region in BeF$_2$ has negative slope,
suggesting that the dynamic crossover in BeF$_2$ upon cooling is related to
the entropy decrease from HDL-like on the high-temperature side to LDL-like
on the low-temperature side.

\section{Summary}

In summary, we systematically study a simple spherically-symmetric
two-scale Jagla potential with both repulsive and attractive parts. We
find a LL phase transition in an accessible region of the P-T phase
diagram. The Jagla potential also displays water-type thermodynamic and
dynamic anomalies, as well as a dynamic crossover which occurs as the
system crosses the Widom line while cooled along constant pressure paths
$P>P_{c}$. Our simulations, similar to simulations of silicon
\cite{sastrynature2003}, show that the dynamics is Arrhenius in the more
ordered phase (HDL for Jagla model) and fragile for the less ordered
phase (LDL for Jagla model). Our study shows that the dynamics is
Arrhenius on the low-temperature side and fragile on the
high-temperature side, as in water.
The dynamic crossover for $P>P_c$ is consistent with (i) the
experimental observation in confined geometries (small pores) of a
fragility transition \cite{chenJCP2004}, and (ii) experimental
observation of a peak in the specific heat upon cooling water at
atmospheric pressure in nanopores \cite{Oguni}.

\subsubsection*{Acknowledgments}

We thank S.-H Chen, D. Chandler, P. G. Debenedetti, G. Franzese,
J. P. Garrahan, P. Kumar, J. M. H. Levelt Sengers, M. Mazza, P. H. Poole,
F. Sciortino, S. Sastry, F. W. Starr, B. Widom, and Z. Yan for helpful
discussions and NSF grant CHE~0096892 for support. We also thank the Boston
University Computation Center for allocation of CPU time. SVB thanks the
Office of the Academic Affairs of Yeshiva University for funding the
high-performace computer cluster.

\end{document}